\documentclass[12pt]{article}
\usepackage{amssymb}
\usepackage{axodraw}
\usepackage{rotate}
\usepackage{epsfig}
\usepackage{psfig}
\usepackage{cite}
\usepackage{graphicx}
\usepackage{wrapfig}
\usepackage{floatflt}

\begin{document}
\setcounter{page}{0}
\thispagestyle{empty}

%#####################################

\newcommand{\eqnzero}{\setcounter{equation}{0}} 
\newcommand{\mpar}[1]{{\marginpar{\hbadness10000%
                      \sloppy\hfuzz10pt\boldmath\bf#1}}%
                      \typeout{marginpar: #1}\ignorespaces}
\def\mnew{\mpar{\hfil NEW \hfil}\ignorespaces}
\newcommand{\bq} {\begin{equation}}
\newcommand{\eq} {\end{equation}}
\newcommand{\bqa}{\begin{eqnarray}}
\newcommand{\eqa}{\end{eqnarray}}
\newcommand{\nll}{\nonumber\\}

\newcommand{\ip }[1]{u\left({#1}        \right)}    % incoming particle
\newcommand{\iap}[1]{{\bar{v}}\left({#1}\right)}    %    "     anti-p
\newcommand{\op }[1]{{\bar{u}}\left({#1}\right)}    % outgoing p
\newcommand{\oap}[1]{v\left({#1}\right)}            %    "     anti-p

\newcommand{\Litwo}{\mbox{${\rm{Li}}_{2}$}}
\newcommand{\alem}{\alpha_{em}}
\newcommand{\alsS}{\alpha^2_{_S}}
\newcommand{\ds }{\displaystyle}
\newcommand{\sss}[1]{\scriptscriptstyle{#1}}
\newcommand{\sla}[1]{/\!\!\!#1}
%----------------  Masses ...
\def\mgn{mgn}
\def\mw {M_{\sss{W}}}
\def\mws{M_{\sss{W}}^2}
\def\mz {M_{\sss{Z}}}
\def\mh {M_{\sss{H}}}
\def\mhs{M_{\sss{H}}^2}
\def\men{m_{\nu_e}}
\def\mel{m_e}
\def\mup{m_u}
\def\mdn{m_d}
\def\mmn{m_{\nu}}
\def\mmo{m_{\mu}}
\def\mch{mch}
\def\mst{mst}
\def\mtn{mtn}
\def\mta{mta}
\def\mtp{m_t}
\def\mbt{m_b}
\def\mf{m_f}
\def\mv{M_{\sss{V}}}
\def\srt{\sqrt{2}}
\def\qel{Q_f}
\def\qmo{Q_f}
\newcommand{\sqrtL}[3]{\sqrt{\lambda\big(#1,#2,#3\big)}}
%---------------------------    Coupling ...
\newcommand{\vpa}[2]{\sigma_{#1}^{#2}}
\newcommand{\vma}[2]{\delta_{#1}^{#2}}
\newcommand{\af}{I^3_f}
\newcommand{\sqs}{\sqrt{s}}
%----------------------------    stw,ctw ...
\newcommand{\stw}{s_{\sss{W}}  }
\newcommand{\ctw}{c_{\sss{W}}  }
\newcommand{\stws}{s^2_{\sss{W}}}
\newcommand{\stwf}{s^4_{\sss{W}}}
\newcommand{\ctws}{c^2_{\sss{W}}}
\newcommand{\ctwf}{c^4_{\sss{W}}}
%-----------------------------   A-B-C-D functions   ------               
\newcommand{\bff}[4]{B_{#1}\big( #2;#3,#4\big)}             
\newcommand{\fbff}[4]{B^{F}_{#1}\big(#2;#3,#4\big)}        
\newcommand{\scff}[1]{C_{#1}}             
\newcommand{\sdff}[1]{D_{#1}}                 
\newcommand{\dffp}[6]{D_{0} \big( #1,#2,#3,#4,#5,#6;}       
\newcommand{\dffm}[4]{#1,#2,#3,#4 \big) }       
%---                  
\newcommand{\tHmus}{\mu^2}
\newcommand{\epsh}{\hat\varepsilon}
\newcommand{\epsb}{\bar\varepsilon}
%-----------------------------------   Equations, Figures, Tables, Sections...

\newcommand{\chapt}[1]{Chapter~\ref{#1}}
\newcommand{\chaptsc}[2]{Chapter~\ref{#1} and \ref{#2}}
\newcommand{\eqn}[1]{Eq.~(\ref{#1})}
\newcommand{\eqns}[2]{Eqs.~(\ref{#1})--(\ref{#2})}
\newcommand{\eqnss}[1]{Eqs.~(\ref{#1})}
\newcommand{\eqnsc}[2]{Eqs.~(\ref{#1}) and (\ref{#2})}
\newcommand{\eqnst}[3]{Eqs.~(\ref{#1}), (\ref{#2}) and (\ref{#3})}
\newcommand{\eqnsf}[4]{Eqs.~(\ref{#1}), 
          (\ref{#2}), (\ref{#3}) and (\ref{#4})}
\newcommand{\eqnsv}[5]{Eqs.(\ref{#1}), 
          (\ref{#2}), (\ref{#3}), (\ref{#4}) and (\ref{#5})}
\newcommand{\tbn}[1]{Table~\ref{#1}}
\newcommand{\tabn}[1]{Tab.~\ref{#1}}
\newcommand{\tbns}[2]{Tabs.~\ref{#1}--\ref{#2}}
\newcommand{\tabns}[2]{Tabs.~\ref{#1}--\ref{#2}}
\newcommand{\tbnsc}[2]{Tabs.~\ref{#1} and \ref{#2}}
\newcommand{\fig}[1]{Fig.~\ref{#1}}
\newcommand{\figs}[2]{Figs.~\ref{#1}--\ref{#2}}
\newcommand{\figsc}[2]{Figs.~\ref{#1} and \ref{#2}}
\newcommand{\sect}[1]{Section~\ref{#1}}
\newcommand{\sects}[2]{Sections~\ref{#1} and \ref{#2}}
\newcommand{\subsect}[1]{Subsection~\ref{#1}}
\newcommand{\appendx}[1]{Appendix~\ref{#1}}

\def\Cmi{c_{-}}
\def\Cpl{c_{+}}
\def\spr{s'}
\def\betap{\beta_{+}}
\def\betam{\beta_{-}}
\def\betapl{\beta^c_{+}}
\def\betami{\beta^c_{-}}
\def\klmi{k^{-}_1}
\def\klpl{k^{+}_1}
\def\betaf{\beta_f}
\def\ph{\phantom{-}}
\def\phph{\phantom{phantomphantom}}
%--------------------------------------------------  end of file: 
%--------------------------------------------------  cpc_add_def.tex
\newcommand {\uu}[1]{\underline{\underline{#1}}}
\newcommand {\dg}[1]{\bf\color{darkgreen}{#1}}
\newcommand {\db}[1]{\bf\color{darkblue}{#1}}
\newcommand {\dbr}[1]{\bf\color{darkbrown}{#1}}
\newcommand {\rbf}{\red \bf}
\def\mz {M_{\sss{Z}}}
\def\mw {M_{\sss{W}}}
\def\mh {M_{\sss{H}}}
\def\mup{m_u}
\def\mdn{m_d}
\def\srt{\sqrt{2}}
\def\order#1{{\mathcal O}\left(#1\right)}
\def\dd{{\mathrm d}}
\def\MSbar{$\overline{\mathrm{MS}}\ $}
\def\GF {G_{\sss F}}
\def\gw {\Gamma_{\sss W}}
\def\bbl{\bf\blue}
\def\itf{I^{(3)}_f}
\def\thmn{\vartheta_{u\nu}}
\def\thmo{\vartheta_{d l}}
\def\thle{\vartheta_{l}}
\def\mml{m_l}
\def\mmf{m_f}

\def\aga{$20^{\circ} < \theta < 160^{\circ}$}
\def\agb{$1^{\circ} < \theta < 179^{\circ}$}
\def\IZ{IZ}
\def\lk{\hspace*{-3mm}}
\def\vmo{v_f}
\def\amo{a_f}
\def\cmi{c_-}
\def\cpl{c_+}
\def\kmi{k_-}
\def\kpl{k_+}
\def\kl{k_1}
\def\kll{k_2}
\def\srll{\sqrt{2}}
\def\vme{v_e}
\def\ame{a_e}
\def\Ts{T^2}
\def\Us{U^2}
\def\Qs{Q^2}
\def\lk{\hspace*{-3mm}}
\def\Nmuu{N_{-}}
\def\Npuu{N_{+}}

\newcommand{\sanc}{{\tt SANC} }
\def\GF {G_{\sss F}}
\def\gw {\Gamma_{\sss W}}
\def\gz {\Gamma_{\sss Z}}
\def\mw {M_{\sss W}}
\def\mz {M_{\sss Z}}
\def\mh {M_{\sss H}}
\def\stw{s_{\sss W}}
\def\ctw{c_{\sss W}}
\newcommand{\GeV}{\unskip\,\mathrm{GeV}}
\newcommand{\MeV}{\unskip\,\mathrm{MeV}}
\newcommand{\hsm}{\hspace*{-1mm}}
\newcommand{\tmW}{{\widetilde{M}}_{\sss W}}
\def\href#1#2{#2}
% #####################################################
$\,$
\vspace*{-1cm}

\begin{flushright}
{\tt IFJPAN-IV-2007-10}\\
{\tt hep-ph/yymmnnn} 
\end{flushright}

\vspace*{\fill}

\begin{center}

   {\LARGE\bf Electroweak radiative corrections to the \\[2mm]

three channels of the process $f_1\bar{f}_1 ZA\to 0$}

\vspace*{1.5cm}
D.~Bardin$^{1}$, S.~Bondarenko$^{1,2}$, L.~Kalinovskaya$^{1}$, G.~Nanava$^{3}$, 
L.~Rumyantsev$^{1}$ and W.~von Schlippe$^{4}$

\vspace*{\fill}

{\normalsize{\it 
$^{1}$ Dzhelepov Laboratory for Nuclear Problems, JINR,        \\
        ul. Joliot-Curie 6, RU-141980 Dubna, Russia;           \\
$^{2}$ Bogoliubov Laboratory of  Theoretical Physics, JINR,    \\ 
        ul. Joliot-Curie 6, RU-141980 Dubna, Russia;           \\
$^{3}$ Institute of Nuclear Physics, PAN, 31-342  Krak\'ow,    \\
        ul. Radzikowskiego 152, Poland,                         \\
on leave from IHEP, TSU, Tbilisi, Georgia;\\
$^{4}$ Theory Division, PNPI RAN, RU-188300 Gatchina, Russia.
}} 
\vspace*{\fill}

\end{center}

\vspace*{\fill}

\begin{abstract}
{We have calculated the electroweak radiative corrections
at the $\cal{O}(\alpha)$ level to the three channels of the
 process $f_1\bar{f}_1 Z A\to 0$ and implemented them into the
 \sanc system.
 Here $A$ stands for the photon and $f_1$ for a first generation fermion
 whose mass is neglected everywhere except
 in arguments of logarithmic functions. The symbol $\to 0$ means
 that 4-momenta of all
 the external particles flow inwards. We present the complete analytical
 results for the covariant and 
 helicity amplitudes  for three cross channels:
 $f_1\bar{f}_1 \to Z\gamma$,  $Z \to  f_1\bar{f}_1 \gamma$ and 
 $f_1 \gamma  \to f_1 Z$. The one-loop scalar form factors of
these channels are simply related by an appropriate permutation of 
 their arguments $s,t,u$. To check the correctness of our
 results we first of all
 observe the independence of the scalar form factors on the gauge
 parameters and 
 the validity of the Ward identity, i.e. external photon transversality, and,
 secondly, compare our numerical results with 
the other independent calculations available to us.
     }
\end{abstract}

\vspace*{\fill}

\centerline{\em To be submitted to EPJC}

\vspace*{\fill}

\newpage

\section{Introduction}
%---------------------
 The group developing the network client-server system \sanc ({\it Support
of Analytic and Numerical calculations for experiments at Colliders}) actively continues 
to implement processes representing an interest for LHC and ILC physics.
 \sanc is one of a few systems including 
 {\rm Feynarts}~\cite{Mertig:1990an,Hahn:1998yk,Hahn:2000kx} and
 {\rm Grace-loop}~\cite{Belanger:2003sd}
in which calculations of elementary particle 
interactions were done at the one-loop precision level. A detailed description
of  version {\tt V.1.00} \sanc was presented in Ref.~\cite{Andonov:2004hi}.
The \sanc client may be downloaded from two \sanc servers Ref.~\cite{homepagesSANC}.

\begin{floatingfigure}{65mm}
\includegraphics[width=6cm]{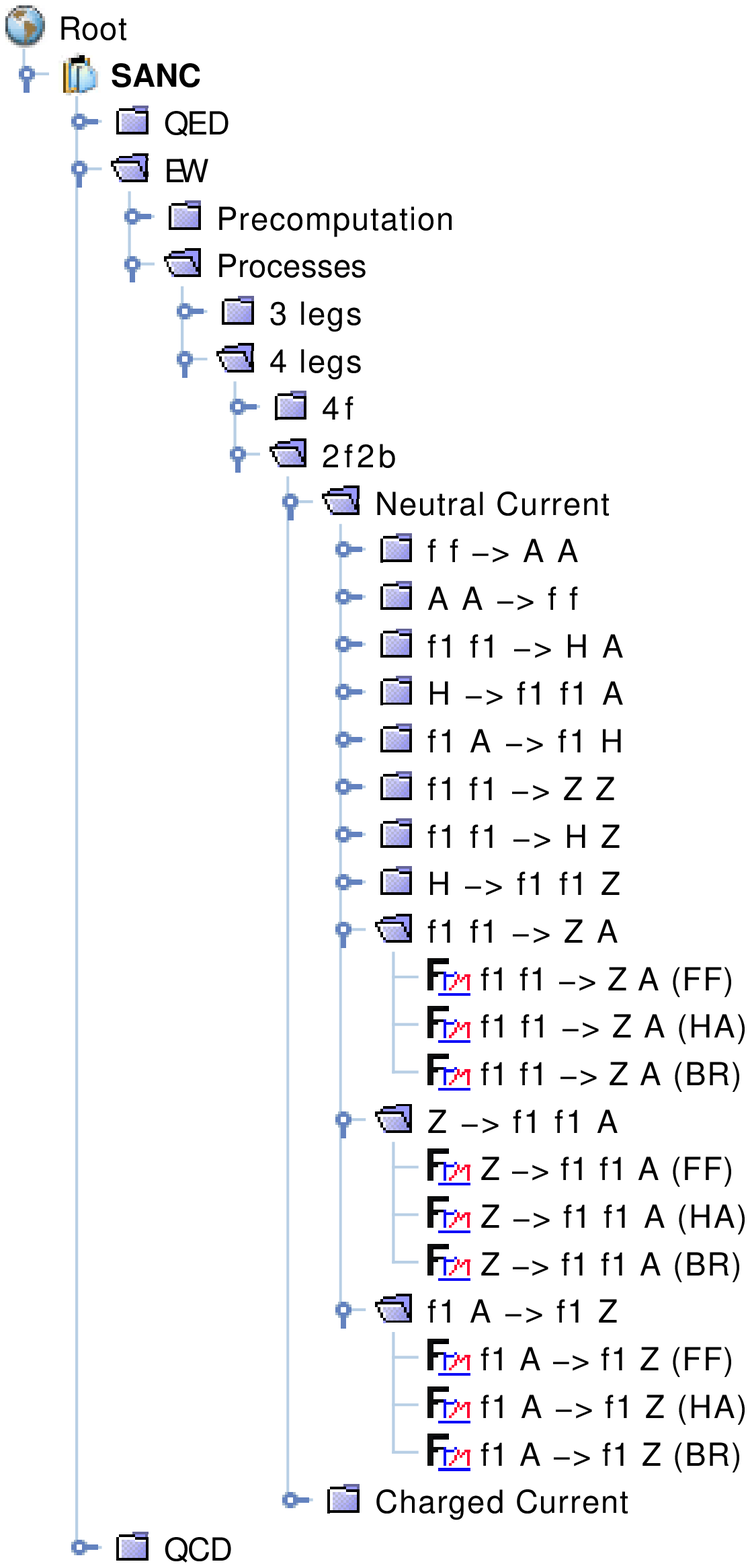}
\caption[New processes in the $ffbb$ sector]
        {New processes in the $ffbb$ sector.}
\label{Processes}
\end{floatingfigure}

 In the recent papers~\cite{Bardin:2007wb,Bardin:2005dp} we presented
 an extension of the
\sanc {\bf Processes} tree in the neutral current $ffbb$ sector,
comprising the version {\tt V.1.10}. In this paper we realize its further extension and
include a calculation of the complete one-loop electroweak radiative corrections to the
$Z$ boson production channels
$f_1\bar{f}_1 \to Z\gamma$ and  $f_1 \gamma  \to f_1 Z$, and 
to the $Z$ boson decay $Z \to  f_1\bar{f}_1 \gamma$.
This class of processes
was already mentioned in section 2.7 of Ref.~\cite{Andonov:2004hi}. For this
reason, we do not change the number of \sanc version, it is still {\tt V.1.10}. 
The new processes are accessible from the $f_1 f_1 \to Z A$, $Z \to f_1 f_1 A$
and $f_1 A \to f_1 Z$ nodes, which are placed in the {\bf Neutral Current} sector of
node {\bf 2f2b} on the electroweak part ({\bf EW}) of the {\bf Processes}
 tree, see Fig.~\ref{Processes}.
Each of these nodes contains standard modules of Scalar Form Factors ({\bf FF}),
Helicity Amplitudes ({\bf HA}), and brem\-sstrahlung ({\bf BR}).

The $Z\gamma$ production process is important for studies of the anomalous
trilinear $Z\gamma\gamma$ and $ZZ\gamma$ gauge boson couplings at the
Fermilab Tevatron\cite{Abazov:2007wy,Abazov:2005ea,Acosta:2004it},
LHC\cite{Baur:2000ae,Haywood:1999qg} and at the Linear
Collider\cite{Atag:2004cn,Walsh:2002gm} in both the $e^+e^-$ and $e\gamma$
modes. The Standard Model (SM) of electroweak interactions predicts no trilinear 
gauge coupling of the Z boson to the photon at the tree level. Any deviation  
of the couplings from the expected values would indicate the existence of new 
physics beyond the SM. At the LHC, it is expected to observe
hundreds of thousands of events of vector boson pair production. To match the
precision of the LHC experiments, the vector boson pair production
processes have to be considered beyond leading order\cite{Accomando:2005ra}.

Leptonic final states of the $Z$ boson decays exhibit a very clear
experimental signature and pave the way for precision tests of the SM beyond the leading
order and possible detection of new physics. That is why it is necessary to fully control higher 
order EW corrections to the fermionic decays of the $Z$ boson.

 These processes were considered in the literature earlier mostly in connection with their
sensitivity to anomalous triple gauge couplings, see for example papers 
~\cite{Rizzo:1996ge,Atag:2003wm,Perez:2004eb,Ananthanarayan:2004eb}.
To our knowledge, the QED and EW corrections to the $Z$ boson production
 have been calculated previously only in papers
~\cite{CapdequiPeyranere:1984sj,Berends:1986yy,Denner:1993kv,Denner:1992qf}.

All the processes under consideration can be treated as various cross channels 
of process  $f_1\bar{f}_1 Z\gamma\to 0$, and hence one-loop corrected scalar form factors, 
derived for this process, can be used for its cross channels also,
after an appropriate permutation of their arguments ($s,t,u$). This is not the case for
helicity amplitudes, however. They are different for all three channels and must be
calculated separately.

The paper is organized as follows.
In section~\ref{Amplitudes} we demonstrate an analytic expression for the
covariant amplitude  at one-loop level in the annihilation channel.
 The helicity amplitudes for all three channels are given  in
 section ~\ref{HAmplitudes}.
In section~\ref{Num_results_comp} we present numerical results
computed by FORTRAN codes generated with software {\tt s2n}
and comparison with other independent calculations.
Finally, summary remarks are given in section~\ref{concl}.

\section{Covariant Amplitude \label{Amplitudes}}
\begin{figure}[!ht]
  \[
   \begin{picture}(125,80)(210,10)
     \GOval(270,40)(34,5)(90){0.02}
     \ArrowLine(240,40)(200,70)
     \ArrowLine(200,10)(240,40)
     \Photon(340,70)(300,40){2}{10}
     %\DashLine(340,70)(300,40){3}
     \Photon(300,40)(340,10){2}{7}

     \ArrowLine(200,16)(220,32)
     \ArrowLine(200,64)(220,48)
     \ArrowLine(340,16)(320,32)
     \ArrowLine(340,64)(320,48)

     \Text(200,54)[]{\large $p_1$}
     \Text(200,28)[]{\large $p_2$}
     \Text(350,28)[]{\large $p_3$}
     \Text(350,54)[]{\large $p_4$}

     \Text(195,80)[]{\large$\bar{f_1}$}
     \Text(195,3)[]{\large $f_1$}
     \Text(350,5)[]{\large $\gamma$}
     \Text(350,80)[]{\large$Z$}
   \end{picture}
  \]
  \vspace*{-5mm}
  \caption [The $\bar{f_1}f_1\gamma Z\to 0$ process]
           {The $\bar{f_1}f_1\gamma Z\to 0$ process.\label{DiagrammaffZACA}}
\end{figure}
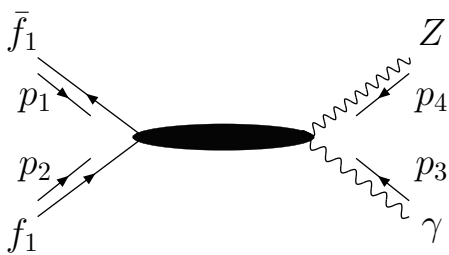

%-----------------------------------------------
 Let us consider the process
\bqa
\bar{f}_1(p_1,\lambda_1) + f_1(p_2,\lambda_2) + \gamma(p_3,\lambda_3) + Z(p_4,\lambda_4) 
\to 0,
\label{decay}
\eqa
where the 4-momenta $p_{i}$ $(i=1,2,3,4)$ of all external particles flow inwards.
 Here, $\lambda_i(i=1,2,3,4)$ are
the helicities of corresponding particles. Schematically this process is
 given  in Fig.~\ref{DiagrammaffZACA}, 
where the black blob represents the sum of all tree and one-loop self energy,
 vertex and box type Feynman diagrams contributing to this process.
 The contributions of the counter term diagrams coming 
from the OMS renormalization procedure is assumed, as well.

%--
%--
We found that  next-to-leading order EW corrections
to this process can be parameterized in terms
of 28 scalar form factors (FF) and corresponding basic matrix elements,
 14 vector and 14 axial ones. For the covariant amplitude (CA) we have:
%---
\bqa
  \label{uniCA}
  {\cal A}_{ \bar{f_1}f_1Z\gamma} =
  \iap{p_1} 
  \left[{\rm Str}^0_{\mu\nu} \left(\vmo {\cal F}^0_{v} + \amo \gamma_5 {\cal F}^0_{a}  \right)
  + \sum_{j=1}^{13}{\rm Str}^j_{\mu\nu} \left({\cal F}^j_{v} +\gamma_5 {\cal F}^j_{a}  \right)
  \right]
  \ip{p_2}\varepsilon^{\gamma}_{\nu}(p_3)\varepsilon^{Z}_{\mu}(p_4),
\eqa
%---
with
\bqa
   {\rm Str}^0_{\mu\nu} &=&
   i \Biggl[ 
     \frac{1}{2} \Biggl(\frac{1}{U^2+m^2_f}+\frac{1}{T^2+m^2_f} \Biggr)
     \gamma_{\mu} \sla{p_3}\gamma_{\nu} 
     +\frac{1}{U^2+m^2_f}  \left(\sla{p_3} \delta_{\mu\nu}-\gamma_{\nu} (p_3)_{\mu}\right)
     \nll &&               
     -\Biggl(\frac{1}{U^2+m^2_f} (p_1)_{\nu}-\frac{1}{T^2+m^2_f} (p_2)_{\nu}\Biggr)\gamma_{\mu} 
     \Biggr],
   \\ 
   {\rm Str}^1_{\mu\nu} &=& i \gamma_{\mu} \sla{p_3}\gamma_{\nu}\,,         
   \nll
   {\rm Str}^2_{\mu\nu} &=&   \sla{p_3}\gamma_{\nu} (p_1)_{\mu}\,,                            
   \nll
   {\rm Str}^3_{\mu\nu} &=&   \sla{p_3}\gamma_{\nu} (p_2)_{\mu}\,,                            
   \nll 
   {\rm Str}^4_{\mu\nu} &=&   \gamma_{\mu} 
   \left[\sla{p_3} (p_1)_{\nu}-\frac{1}{2}\left(U^2+m^2_f\right)\gamma_{\nu} \right],
   \nll 
   {\rm Str}^5_{\mu\nu} &=&  \gamma_{\mu} 
   \left[\sla{p_3} (p_2)_{\nu}-\frac{1}{2}\left(T^2+m^2_f\right)\gamma_{\nu} \right],
   \nll
   {\rm Str}^6_{\mu\nu} &=& i 
   \left[\sla{p_3} (p_1)_{\nu}-\frac{1}{2}\left(U^2+m^2_f\right)\gamma_{\nu} \right] (p_1)_{\mu},
   \nll
   {\rm Str}^7_{\mu\nu} &=& i
   \left[\sla{p_3} (p_2)_{\nu}-\frac{1}{2}\left(T^2+m^2_f\right)\gamma_{\nu} \right] (p_1)_{\mu},
   \nll 
   {\rm Str}^8_{\mu\nu} &=& i
   \left[\sla{p_3} (p_1)_{\nu}-\frac{1}{2}\left(U^2+m^2_f\right)\gamma_{\nu} \right] (p_2)_{\mu},
   \nll
   {\rm Str}^9_{\mu\nu} &=& i
   \left[\sla{p_3} (p_2)_{\nu}-\frac{1}{2}\left(T^2+m^2_f\right)\gamma_{\nu} \right] (p_2)_{\mu},
   \nll 
   {\rm Str}^{10}_{\mu\nu} &=& i
   \left(\sla{p_3} \delta_{\mu \nu}-\gamma_{\nu} (p_3)_{\mu} \right),   
   \nll         
   {\rm Str}^{11}_{\mu\nu} &=& i \gamma_{\mu} 
   \biggl[\left(T^2+m^2_f\right) (p_1)_{\nu}-\left(U^2+m^2_f\right) (p_2)_{\nu}\biggr],  
   \nll  
   {\rm Str}_{12} &=& 
   (p_1)_{\mu} (p_2)_{\nu}+(p_2)_{\mu} (p_2)_{\nu}+\frac{1}{2}\left(T^2+m^2_f\right)\delta_{\mu\nu},
   \nll 
   {\rm Str}^{13}_{\mu\nu} &=&
   \biggl[\left(T^2+m^2_f\right) (p_1)_{\nu}-\left(U^2+m^2_f\right) (p_2)_{\nu}\biggr] (p_2)_{\mu},
   \nonumber
\eqa
%--
where $\iap{p_1}$, $\ip{p_2}$ and $m_f$ are the bispinors and the mass of
 the external fermions, respectively; 
$\varepsilon^{\gamma}_{\nu}(p_3)$ denotes the photon polarization vector
and $\varepsilon^{Z}_{\mu}(p_4)$ is the $Z$ boson polarization
vector; the vector and axial gauge-boson-to-fermion couplings are 
denoted by $v_f$ and $a_f$, respectively; ${\cal F}^j_{v,a}$ are the scalar FF
of the vector and axial vector currents, respectively;
${\cal F}^0_{v,a}$ and ${\rm Str}^{0}_{\mu\nu}$
correspond to the lowest-order matrix elements.
The usual Mandelstam invariants in Pauli metric ($p^2=-m^2$) are defined as
follows:
%--
\bqa
&& (p_1+p_2)^2=Q^2=-s,
\nll
&& (p_2+p_3)^2=T^2=-t,
\nll
&& (p_2+p_4)^2=U^2=-u.
\eqa
%--       
In Eq.(\ref{uniCA}) we keep the fermion mass in order to maintain photon transversality.
Moreover in mass-containing denominators of ${\rm Str}^{0}_{\mu\nu}$, the mass
cannot be neglected because these denominators correspond to the propagators of fermions 
which emit external photons and thus would lead to mass singularities.

The basic matrix elements, ${\rm Str}^{j}_{\mu\nu}$, are chosen to be explicitly
transverse to the photonic 4-momentum. 
That is, for all of them the following relations hold:
%--
\bqa
    {\rm Str}^{j}_{\mu\nu} (p_3)_{\nu} = 0.
\eqa
%--
We have checked that the FF ${\cal F}^j_{v,a}$ are free of gauge parameters
and of ultraviolet singularities (all calculations are done in the $R_\xi$ gauge).
The analytical expressions of the FF are too cumbersome to be presented in this paper.
They can be reproduced on-line with help of the \sanc system.
The CA for the processes we are interested in can be obtained
from Eq.(\ref{uniCA}) exploiting crossing symmetry. This subject is covered
in the next section.

\section{Helicity Amplitudes \label{HAmplitudes}}
%------------------------------------------------
 In this section we collect the analytical expressions of the helicity
amplitudes (HA) for all three channels. Let us briefly recall the \sanc strategy
of observable (cross section, differential distributions) calculations.
In a first step, \sanc constructs the CA of the process,
free of gauge parameters and of ultraviolet singularities,
taking into account  all lowest order
and one-loop Feynman diagrams that contribute to the process.
In the next step, HA are calculated analytically and
converted into numerical code. Further, the cross section or the decay width of
the process is formed as the incoherent sum of squares of all possible HA:
%--
\bqa
 d\sigma(d\Gamma) \sim \sum_{\lambda_1\lambda_2...\lambda_n}
|{\cal H}_{\lambda_1\lambda_2...\lambda_n}|^2\, d\Phi^{n}
\eqa 
%--
where squaring and summing is performed numerically. 
And finally, the Monte Carlo integrations over phase-space $d\Phi^{n}$ are performed using
Vegas routine~\cite{Lepage:1977sw}.

\subsection{Annihilation channel $\bar{f}_1f_1\to Z\gamma $}
%-----------------------------------------------------------
  To obtain the CA for the process
%-
 \bqa
  \bar{f}_1(p_1,\lambda_1) + f_1(p_2,\lambda_2) \to \gamma(p_3,\lambda_3) + Z(p_4,\lambda_4),
    \label{annihil}
 \eqa
%-
\noindent where $\lambda_i(i=1,2,3,4)$ are the helicities of the external particles, 
 we use the following substitutions of 4-momenta in Eq.(\ref{uniCA}):
\[
  \begin{array}{llll}
    & p_1 & \to &~~p_1, \\
    & p_2 & \to &~~p_2, \\
    & p_3 & \to & -p_3, \\
    & p_4 & \to & -p_4. \\
  \end{array}
\]

\vskip 2mm
\begin{figure}[!ht]
   \[
    \begin{picture}(125,80)(210,-20)
    %\begin{picture}(145,80)(230,-20)
    \GOval(270,40)(34,5)(90){0.02}
    \ArrowLine(240,40)(200,70)
    \ArrowLine(200,10)(240,40)
    %\DashLine(340,70)(300,40){3}
    \Photon(340,70)(300,40){3}{10}
    \Photon(300,40)(340,10){2}{7}

    \ArrowLine(200,16)(220,32)
    \ArrowLine(200,64)(220,48)
    \ArrowLine(320,32)(340,16)
    \ArrowLine(320,48)(340,64)
    
    \Text(200,54)[]{\large $p_1$}
    \Text(200,28)[]{\large $p_2$}
    \Text(350,28)[]{\large $p_3$}
    \Text(350,54)[]{\large $p_4$}
    
    \Text(200,80)[]{\large$\bar{f_1}$}
    \Text(200,3)[]{\large $f_1$}
    \Text(350,5)[]{\large $\gamma$}
    \Text(350,80)[]{\large$Z$}
      \end{picture}
   \]
   \vspace*{-16mm}
   \caption [Schematic representation of one-loop Feynman diagrams for the annihilation channel]
            {Schematic representation of one-loop Feynman diagrams for the annihilation channel.
            \label{DiagrammaffHA}}
\end{figure}
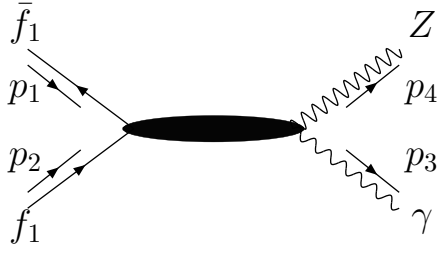

 The set of non-vanishing HA for this process, which we denote as 
 ${\cal H}_{\lambda_1\lambda_2\lambda_3\lambda_4}$, read:
%--
\bqa
%1
{\cal H}_{\mp\mp\mp\mp} &=& \frac{\mf}{\sqrt{s}} 
\left[ 2-\frac{1}{2}\frac{s Z_4(M_Z)}{Z_1(\mf)Z_2(\mf)}\sin^2\vartheta_\gamma\right]\vmo{\cal F}_{v0}
\nll &&
+ \frac{Z_4(\mz)}{4\sqrt{s}}\cmi\Biggl[
  \cpl \Bigl({\cal F}_{2}^{\pm} -  {\cal F}_{3}^{\pm} - {\cal F}_{4}^{\pm} \Bigr)
      - \cmi {\cal F}_{5}^{\pm} +  {\cal F}_{12}^{\pm} - \frac{s}{2}\cpl {\cal F}_{13}^{\pm} 
                                \Biggr],
\nll
%2
{\cal H}_{\mp\mp\mp0} &=&\mp \frac{Z_4(\mz)}{4\srll\mz}\sin\vartheta_\gamma\Biggl[ 
         \frac{4\mz^2\mf}{Z_1(\mf) Z_2(\mf)}\cos\vartheta_\gamma \vmo {\cal F}_{v0}
\nll &&
   + \kll{\cal F}_{2}^{\pm} + \kl{\cal F}_{3}^{\pm} - \kll{\cal F}_{4}^{\pm}
   + \kpl\cmi{\cal F}_{5}^{\pm} - s{\cal F}_{12}^{\pm} + \frac{s}{2}\kl{\cal F}_{13}^{\pm}
                                                                        \Biggr], 
\nll
%3
{\cal H}_{\mp\mp\mp\pm} &=& \frac{\sqrt{s}}{4} Z_4(\mz)\sin^2\vartheta_\gamma \Biggl[ 
         \frac{2 \mf}{Z_1(\mf) Z_2(\mf)} \vmo{\cal F}_{v0}
 - {\cal F}_{2}^{\pm} + {\cal F}_{3}^{\pm} + {\cal F}_{4}^{\pm} - {\cal F}_{5}^{\pm}
            + \frac{s}{2} {\cal F}_{13}^{\pm}\Biggr], 
\nll
%4                                               
{\cal H}_{\mp\mp\pm\mp} &=& \frac{\sqrt{s}}{8} Z_4(\mz)\sin^2\vartheta_\gamma \Biggl[
 \frac{4\mf}{Z_1(\mf) Z_2(\mf)}\vmo{\cal F}_{v0} + s {\cal F}_{13}^{\pm}\Biggr], 
\nll
%5
{\cal H}_{\mp\mp\pm 0} &=& \pm\frac{Z_4(\mz)}{\srll\mz}\sin\vartheta_\gamma\Biggl[
        \frac{\mf}{Z_1(\mf)Z_2(\mf)}
\left( \mz^2 \cos\vartheta_\gamma \vmo{\cal F}_{v0}\pm Z_4(\mz)\amo{\cal F}_{a0}\right)
\nll &&
  -\frac{s}{4}\Bigl(2{\cal F}_{4}^{\pm}+{\cal F}_{12}^{\pm}-\frac{1}{2}\kl{\cal F}_{13}^{\pm}\Bigr)
                                                                       \Biggr],
\nll
%6
{\cal H}_{\mp\mp\pm\pm} &=& \frac{\mf}{2\sqrt{s}} \Biggl[
\left(4-\left(2 Z_4(\mz)+s\sin^2\vartheta_\gamma\right)
      \frac{Z_4(\mz)}{Z_1(\mf)Z_2(\mf)}\right)\vmo {\cal F}_{v0}
\nll&&
 \pm 2\frac{Z^2_4(\mz)}{Z_1(\mf)Z_2(\mf)}\cos\vartheta_\gamma\amo {\cal F}_{a0}
                                                   \Biggr]
%\nll&&
- \frac{\sqrt{s}}{2}Z_4(\mz)\left(
    \cpl{\cal F}_{4}^{\pm}-\frac{1}{2}\cmi{\cal F}_{12}^{\pm}+\frac{s}{4}
 \sin^2\vartheta_\gamma{\cal F}_{13}^{\pm}
                             \right), 
\nll
%7
{\cal H}_{\pm\mp\pm\pm} &=&\mp \frac{1}{8}\sin\vartheta_\gamma \Biggl[
  \frac{4\mz^2}{Z_1(\mf)}{\cal F}_{0}^{\pm}
        - Z_4(\mz)\left[s\cpl\left({\cal F}_{6}^{\pm}-{\cal F}_{8}^{\pm}\right)
        + 4{\cal F}_{10}^{\pm} + 2s\cmi{\cal F}_{11}^{\pm} \right] \Biggr],
\nll
%8
{\cal H}_{\pm\mp\mp\mp} &=&\pm\frac{1}{8}\sin\vartheta_\gamma\Biggl[
   \frac{4\mz^2}{Z_2(\mf)}{\cal F}_{0}^{\pm}
     - Z_4(\mz) \left[8{\cal F}_{1}^{\pm}
     + s\cmi\left({\cal F}_{7}^{\pm}-{\cal F}_{9}^{\pm}\right)
     - 4 {\cal F}_{10}^{\pm} + 2 s\cpl{\cal F}_{11}^{\pm} \right] \Biggr],
\nll
%9
{\cal H}_{\pm\mp\pm 0} &=& \frac{1}{8\srll}\frac{\sqrt{s}}{\mz}\cpl\Biggl[
    \frac{8\mz^2}{Z_1(\mf)}{\cal F}_{0}^{\pm}
      + Z_4(\mz)\left( \kll{\cal F}_{6}^{\pm}
                      + \kl{\cal F}_{8}^{\pm}
                       - 4 {\cal F}_{10}^{\pm}
                - 2\kpl\cmi{\cal F}_{11}^{\pm}\right) \Biggr],
\nll
%10
{\cal H}_{\mp\pm\pm 0} &=& -\frac{1}{8\srll}\frac{\sqrt{s}}{\mz}\cmi\Biggl[  
  \frac{8\mz^2}{Z_2(\mf)}{\cal F}_{0}^{\pm}
      - Z_4(\mz)\left(   8 {\cal F}_{1}^{\pm}
                     + \kll{\cal F}_{7}^{\pm}
                     + \kl {\cal F}_{9}^{\pm}
                     -   4 {\cal F}_{10}^{\pm}
                + 2\kpl\cpl{\cal F}_{11}^{\pm}
                                 \right)\Biggr],
\nll
%11
{\cal H}_{\pm\mp\pm\mp} &=&\mp \frac{s}{8} Z_4(\mz) \sin\vartheta_\gamma \cpl \Biggl[ 
 \frac{2}{Z_1(\mf)Z_2(\mf)}{\cal F}_{0}^{\pm}
                         + {\cal F}_{6}^{\pm}
                         - {\cal F}_{8}^{\pm}
                       - 2 {\cal F}_{11}^{\pm}\Biggr], 
\nll
%12
{\cal H}_{\pm\mp\mp\pm} &=&\pm \frac{s}{8} Z_4(\mz) \sin\vartheta_\gamma \cmi \Biggl[ 
 \frac{2}{Z_1(\mf)Z_2(\mf)}{\cal F}_{0}^{\pm}
                         + {\cal F}_{7}^{\pm}
                         - {\cal F}_{9}^{\pm}
                       - 2 {\cal F}_{11}^{\pm}\Biggr],
\eqa
%---
\vskip 2mm
\noindent with the following shorthand notation
%---
\bqa
  {\cal F}^{\pm}_{0}&=&\vmo{\cal F}_{v0}(s,t,u)\pm\amo{\cal F}_{a0}(s,t,u),
  \nll
  {\cal F}^{\pm}_{j}&=&{\cal F}_{vj}(s,t,u)\pm{\cal F}_{aj}(s,t,u),\qquad j=1,...13,
  \nll
  k_{1,2}&=&s c_{\pm}-\mz^2 c_{\mp}\,,\;\;\;c_{\pm}=1\pm\cos\vartheta_\gamma\,,
  \nll
  Z_1(\mf)&=&\frac{1}{2}Z_4(\mz)\left(1+\beta\cos{\vartheta_{\gamma}}\right), 
\quad \beta\equiv\beta_f=\sqrt{1-4m^2_f/s},
  \nll
  Z_2(\mf)&=&\frac{1}{2}Z_4(\mz)\left(1-\beta\cos{\vartheta_{\gamma}}\right),
  \nll
  Z_4(\mz)&=&s-\mz^2.
  \label{anndef}
\eqa
%---

\noindent
Here $\vartheta_{\gamma}$ is the center of mass system angle of the produced photon 
(angle between momenta $\vec{p}_2$ and $\vec{p}_3$), $t$ and $u$ are the Mandelstam
variables:
%---
\bqa
  t=\mf^2-Z_2(\mf), \qquad u=\mf^2-Z_1(\mf).
  \label{anndef1}
\eqa
 %---

\subsection{Decay channel $Z\to f_1\bar{f}_1\gamma$}
%---------------------------------------------------
The CA of $Z$ boson decay into fermion anti-fermion pairs and one real photon,
%---
\bqa
  Z(p_2,\lambda_2) \to \gamma(p_1,\lambda_1) + f_1(p_3,\lambda_3) + \bar{f}_1(p_4,\lambda_4), 
 \label{decay1}
\eqa
%---
 is obtained by interchanging of 4-momenta in Eq.(\ref{uniCA}) as follows:
 
\[
 \begin{array}{llll}
    &\lk p_1 & \to & -p_3,   \\
    &\lk p_2 & \to & -p_4,   \\
    &\lk p_3 & \to & -p_1,   \\
    &\lk p_4 & \to & ~~p_2.
 \end{array}
\]

\begin{figure}[!ht]
  \[
  \begin{picture}(125,80)(230,20)
   \Photon(210,70)(260,70){3}{10}
   \Photon(260,70)(300,100){2}{7}
   \Vertex(260,70){1.5}
   \GOval(280,55)(30,5)(54){0.02}
   \ArrowLine(340,70)(300,40)
   \ArrowLine(300,40)(340,10)
   \ArrowLine(320,32)(340,16)
   \ArrowLine(320,48)(340,64)
   \ArrowLine(270,85)(290,101)
   \ArrowLine(220,75)(245,75)
   \Text(205,70)[]{\large $Z$}
   \Text(305,105)[]{\large $\gamma$}
   \Text(350,10)[]{\large $f_1$}
   \Text(350,75)[]{\large $\bar{f}_1$}
   \Text(235,85)[]{\large $p_2$}
   \Text(270,100)[]{\large $p_1$}
   \Text(350,28)[]{\large $p_3$}
   \Text(350,54)[]{\large $p_4$}
 \end{picture}
 \]
%\vspace*{-5mm}
 \caption [Schematic representation of one-loop Feynman diagrams for the decay channel]
          {Schematic representation of one-loop Feynman diagrams for the decay channel.}
          \label{DiagrammaHAff}
\end{figure}
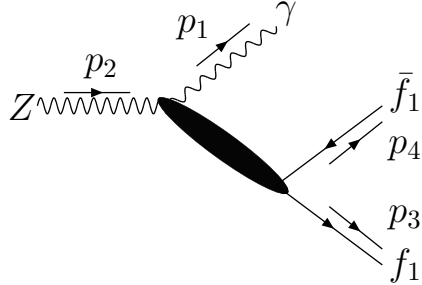

\noindent For the non-vanishing HA, ${\cal H}_{\lambda_2\lambda_1\lambda_3\lambda_4}$, we have
% 1
\bqa
{\cal H}_{\pm\pm\pm\pm} &=& -\frac{\sqrt{s}}{8} Z_2(\mz)\sin^2\vartheta_{f} 
            \Biggl[ \frac{4 \mf}{ Z_3(\mf)Z_4(\mf)} \vmo {\cal F}_{v0}
                    + s {\cal F}_{13}^{\pm} \Biggr],
\nll
% 2
{\cal H}_{\mp\mp\mp\pm} &=& \mp\frac{s}{8} Z_2(\mz) \sin\vartheta_{f}\cpl \Biggl[
\frac{2}{ Z_3(\mf) Z_4(\mf)} {\cal F}^{\pm}_{0} 
 + {\cal F}_{7}^{\pm} - {\cal F}_{9}^{\pm} - 2 {\cal F}_{11}^{\pm} \Biggr],
\nll
% 3
{\cal H}_{\pm\pm\mp\pm} &=& \pm\frac{s}{8} Z_2(\mz) \sin\vartheta_{f}\cmi \Biggl[
\frac{2}{Z_3(\mf) Z_4(\mf)}{\cal F}^{\pm}_{0}
 + {\cal F}_{6}^{\pm} - {\cal F}_{8}^{\pm}  - 2 {\cal F}_{11}^{\pm} \Biggr],
\nll
% 4
{\cal H}_{\mp\pm\mp\pm} &=& \mp\frac{1}{8} \sin\vartheta_{f} \Biggl[
 4\frac{\mz^2}{Z_3(\mf)}{\cal F}^{\pm}_{0}
 - Z_2(\mz) \Biggl( s\cmi \left( {\cal F}_{6}^{\pm} - {\cal F}_{8}^{\pm} \right)
+ 4 {\cal F}_{10}^{\pm} + 2 s\cpl {\cal F}_{11}^{\pm} \Biggr)\Biggr],
\nll
% 5
{\cal H}_{\pm\mp\mp\pm} &=& \pm\frac{1}{8} \sin\vartheta_{f} \Biggl[
 4\frac{\mz^2}{Z_4(\mf)}{\cal F}^{\pm}_{0}
 - Z_2(\mz) \Biggl(8 {\cal F}_{1}^{\pm}
 + s\cpl\left({\cal F}_{7}^{\pm} - {\cal F}_{9}^{\pm} \right)
 - 4{\cal F}_{10}^{\pm} + 2 s\cmi {\cal F}_{11}^{\pm} \Biggr)\Biggr],
\nll
% 6
{\cal H}_{\mp\mp\pm\pm} &=& -\frac{1}{8}\sqrt{s} Z_2(\mz) \sin^2\vartheta_f \Biggl[
\frac{4\mf}{ Z_3(\mf) Z_4(\mf)}\vmo {\cal F}_{v0}
                   - 2\left(  {\cal F}_{2}^{\pm} 
                             -{\cal F}_{3}^{\pm} 
                             -{\cal F}_{4}^{\pm} 
                            + {\cal F}_{5}^{\pm}  \right)
                          + s {\cal F}_{13}^{\pm} \Biggr],
\nll
% 7
{\cal H}_{\pm\mp\pm\pm} &=&-\frac{1}{4}\sqrt{s} \Biggl[
 8\frac{\mf}{s}\frac{\mz^2}{Z_2(\mz)}\vmo {\cal F}_{v0}
\nll &&\hspace*{10mm}
 - Z_2(\mz)\Biggl(\sin^2\vartheta_f\Bigl({\cal F}_2^{\pm} - {\cal F}_3^{\pm} - {\cal F}_4^{\pm}\Bigr)
- \cpl^2 {\cal F}_5^{\pm} + \cpl {\cal F}_{12}^{\pm}   
        + \frac{s}{2}\sin^2\vartheta_f{\cal F}_{13}^{\pm} \Biggr) \Biggr],
\nll
% 8                                                     
{\cal H}_{\mp\pm\pm\pm} &=&-\frac{1}{2}\sqrt{s} \Biggl[
  4\frac{\mf}{s}\frac{\mz^2}{Z_2(\mz)}\vmo{\cal F}_{v0}
 -2\frac{\mf}{s}\frac{Z^2_2(\mz)}{Z_3(\mf)Z_4(\mf)}\left(\vmo{\cal F}_{v0}
                                          \pm \cos\vartheta_f\amo{\cal F}_{a0}\right)
\nll &&\hspace*{10mm}
+ Z_2(\mz)\Biggl(\cmi {\cal F}_4^{\pm} - \frac{1}{2}\cpl {\cal F}_{12}^{\pm} 
+ \frac{s}{4}\sin^2\vartheta_f {\cal F}_{13}^{\pm} \Biggr) \Biggr],  
\nll
% 9
{\cal H}_{0\pm\pm\pm} &=& \frac{i}{8\srll} \frac{s Z_2(\mz)}{\mz} \sin\vartheta_f \Biggl[
    \frac{8\mf}{s Z_3(\mf) Z_4(\mf) }
    \left(\mz^2 \cos\vartheta_f \vmo {\cal F}_{v0} \pm {Z_2(\mz)} \amo {\cal F}_{a0} \right)
\nll &&\hspace*{10mm}
   + 4 {\cal F}_{4}^{\pm} + 2 {\cal F}_{12}^{\pm}
                         - \kll {\cal F}_{13}^{\pm} \Biggl],
\nll
% 10
{\cal H}_{0\mp\pm\pm} &=& \frac{i}{4\srll} \frac{Z_2(\mz)}{\mz} \sin\vartheta_f \Biggl[
  \frac{4s\mf}{Z_3(\mf)Z_4(\mf)}\cos\vartheta_f\vmo{\cal F}_{v0}
\nll &&\hspace*{10mm}
 - \kl {\cal F}_2^{\pm}  - \kll {\cal F}_3^{\pm} + \kl {\cal F}_4^{\pm} - \kpl\cpl {\cal F}_5^{\pm}   
 + s {\cal F}_{12}^{\pm} - \frac{s}{2}\kll {\cal F}_{13}^{\pm}\Biggl],
\nll
% 11
{\cal H}_{0\mp\mp\pm} &=&\mp\frac{i}{\srll}\frac{\sqrt{s}}{\mz} \Biggl[
 2\frac{\mz^2}{Z_2(\mz)}{\cal F}^{\pm}_{0}
 -\frac{1}{8}Z_2(\mz)\cpl\Biggl(8 {\cal F}_1^{\pm}
 + \kl {\cal F}_7^{\pm} + \kll{\cal F}_9^{\pm} - 4 {\cal F}_{10}^{\pm}
 + 2\kpl\cmi {\cal F}_{11}^{\pm} \Biggr) \Biggr],
\nll
% 12
{\cal H}_{0\pm\mp\pm} &=&\mp\frac{i}{8\srll}\frac{\sqrt{s}}{\mz} \Biggl[
 2\frac{\mz^2}{Z_2(\mz)}{\cal F}^{\pm}_{0}
 +\frac{1}{8}Z_2(\mz)\cmi\Biggl(
 \kl {\cal F}_{6}^{\pm} + \kll{\cal F}_{8}^{\pm} - 4 {\cal F}_{10}^{\pm}  
 - 2\kpl\cpl {\cal F}_{11}^{\pm} \Biggl) \Biggr],
\eqa
%---
\noindent
where ${\cal F}^{\pm}_j$ and the coefficients $k_{1,2}$ are defined by 
Eqs.~(\ref{anndef}) and (\ref{anndef1}) with $c_{\pm}=1\pm\cos\vartheta_f$, and
%---
\bqa
  && Z_3(\mf)=\frac{1}{2}Z_2(\mz)\left(1+\beta\cos{\vartheta_{f}}\right),
\qquad \beta=\sqrt{1-4m_f^2/s},
  \nll
  &&Z_4(\mf)=\frac{1}{2}Z_2(\mz)\left(1-\beta\cos{\vartheta_{f}}\right),
  \nll
  && s=M^2_{f\bar{f}}, \quad t=\mf^2+Z_4(\mf), \quad u=\mf^2+Z_3(\mf).
\eqa
%---
\noindent Here $Z_2(\mz)=\mz^2-s$ and $\vartheta_{f}$ is the angle between 
the vector $\vec{p}_3$ and the direction defined by the photon momentum $\vec{p}_1$
in the rest frame of compound $(\vec{p}_3,\vec{p}_4)$.
The photon momentum, $\vec{p}_1$, is chosen to be direction of the $z$-axes in
the $(\vec{p}_3,\vec{p}_4)$ rest frame.

\subsection{$Z$ production channel $e \gamma \to e Z$}
%----------------------------------------------------- 
 And finally, in order to obtain the CA for the $Z$ boson production channel
%---
\bqa
  \gamma(p_1,\lambda_1) + e^{\pm}(p_2,\lambda_2) \to e^{\pm}(p_3,\lambda_3) + Z(p_4,\lambda_4)
\eqa
%---
from  Eq.(\ref{uniCA}), the 4-momenta permutations  must be chosen as follows:
 \[
  \begin{array}{llll}
    &\lk p_1 & \to & -p_3,   \\
    &\lk p_2 & \to & -p_4,   \\
    &\lk p_3 & \to & -p_1,   \\
    &\lk p_4 & \to & ~~p_2.
  \end{array}
 \]

\vskip 5mm
\begin{figure}[!ht]
  \[
  \begin{picture}(125,80)(212,0)
    \GOval(270,40)(34,5)(0){0.02}
    \Photon(270,67)(230,97){2}{7}
    \Photon(310,97)(270,67){3}{10}
    \ArrowLine(230,-17)(270,13)
    \ArrowLine(270,13)(310,-17)
    \ArrowLine(240,-20)(260,-4)
    \ArrowLine(240,100)(260,84)
    \ArrowLine(280,84)(300,100)
    \ArrowLine(280,-4)(300,-20)
    \Text(230,80)[]{\large $\gamma$}
    \Text(230,5)[]{\large $e$}
    \Text(310,5)[]{\large $e$}
    \Text(310,80)[]{\large $Z$}
    \Text(260,100)[]{\large $p_1$}
    \Text(285,100)[]{\large $p_4$}
    \Text(285,-20)[]{\large $p_3$}
    \Text(260,-20)[]{\large $p_2$}
  \end{picture}
  \]
  \vspace*{1mm}
  \caption [Schematic representation of one-loop Feynman diagrams for the $Z$ boson production channel]
           {Schematic representation of one-loop Feynman diagrams for the $Z$ boson production channel.}
           \label{DiagrammaAffH}
\end{figure}
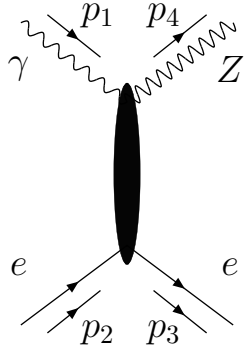

 The HA, ${\cal H}_{\lambda_1\lambda_2\lambda_3\lambda_4}$, for this channel read 
%1
\bqa
{\cal H}_{\pm\mp\mp\pm} &=&  \frac{k_3}{\sqrt{2 s} }\Biggl[ 
  2\left(\frac{1}{\kmi}-\frac{1}{Z_3(\mel)}\right)\mz^2 {\cal F}_0^{\pm}
  + s \cmi\left(\frac{\kmi}{4}\cpl{\cal F}_8^{\pm}
                                - {\cal F}_{10}^{\pm}
                            + \kmi{\cal F}_{11}^{\pm} \right) \Biggr],
\nll
%2 
{\cal H}_{\pm\mp\mp 0} &=& \pm \frac{k_4}{\mz} \Biggl[ 
          \frac{2\mz^2}{\kmi}{\cal F}_0^{\pm}
  +\frac{1}{2}\cpl \left(
             \frac{\kmi^2}{2}{\cal F}_{6}^{\pm}   
         -\frac{\kmi}{4}  \kl{\cal F}_{8}^{\pm}   
                       + \kpl{\cal F}_{10}^{\pm}
                     - s \kmi{\cal F}_{11}^{\pm} \right) \Biggr],
\nll 
%3
{\cal H}_{\pm\mp\pm 0} &=& \frac{k_3}{\mz \sqrt{s}} \Biggl[
\frac{2\mel}{Z_3(\mel)}
  \Biggl(\mz^2 \left(\frac{s}{\kmi}+1\right) \vme {\cal F}_{v_0} 
                              \pm \frac{s^2}{\kmi} \ame {\cal F}_{a_0}\Biggr)
\nll &&
      +\frac{s}{4}  \cmi \left(
                     2 \kmi{\cal F}_{4}^{\pm}  
                     + \kpl{\cal F}_{12}^{\pm}
       -\frac{\kmi}{2} \kl {\cal F}_{13}^{\pm} \right) \Biggr],
\nll \
%4
{\cal H}_{\pm\mp\pm\pm} &=&
  \mp \frac{k_4}{\srll}  \Biggl[ 
    \frac{2 s \mel}{Z_3(\mel) \kmi} {\cal F}_{v_0}^{\pm}
  + \frac{s}{2} \cmi \left(
                         {\cal F}_{12}^{\pm}
   - \frac{\kmi}{2}\cpl {\cal F}_{13}^{\pm} \right)\Biggr],
\nll
%5 
{\cal H}_{\pm\mp\pm\mp} &=& \mp \frac{k_4}{\srll}\cpl  \Biggl[  \kmi {\cal F}_{4}^{\pm}
         + \frac{s}{2} \left( {\cal F}_{12}^{\pm}
      + \frac{\kmi}{2}  \cmi{\cal F}_{13}^{\pm} \right)\Biggr],
\nll
%6                                                                               
{\cal H}_{\mp\mp\pm 0} &=&-\frac{\sqrt{s} k_3}{\mz} \Biggl[ 
  \frac{2 \mz^2 \mel}{s Z_3(\mel)} \vme  {\cal F}_{v_0}
\nll &&
                                  - \kmi {\cal F}_{2}^{\pm}
                        + \frac{1}{2}\kl {\cal F}_{3}^{\pm}
                                  + \kmi {\cal F}_{4}^{\pm}
                                     - s {\cal F}_{5}^{\pm}
     + \frac{1}{4} \cmi\left(      \kpl  {\cal F}_{12}^{\pm}
                  - \frac{\kmi}{2}   \kl {\cal F}_{13}^{\pm} \right)\Biggr],
\nll
%7
{\cal H}_{\mp\mp\mp\mp} &=&
 \left(\frac{s}{2}\right)^\frac{3}{2}  k_3 \cmi   {\cal F}_{9}^{\pm},
\nll
%8
{\cal H}_{\mp\mp\mp 0} &=& \pm \frac{s k_4}{\mz}   \Biggl[
                             2 {\cal F}_{1}^{\pm}
             - \frac{\kmi}{2}  {\cal F}_{7}^{\pm}
              + \frac{1}{4}\kl {\cal F}_{9}^{\pm}
                             - {\cal F}_{10}^{\pm}
         + \frac{\kmi}{2} \cpl {\cal F}_{11}^{\pm} \Biggr],                                               
\nll 
%9
{\cal H}_{\mp\mp\mp\pm} &=&-\frac{k_3}{ \sqrt{2 s}} \Biggl[
           \frac{2\kmi}{Z_3(\mel)}  {\cal F}_{0}^{\pm}
                - s \left( 4 {\cal F}_{1}^{\pm}
         - \frac{s}{2}  \cmi {\cal F}_{9}^{\pm}
                          -2 {\cal F}_{10}^{\pm}
                 - \kmi \cmi {\cal F}_{11}^{\pm} \right) \Biggr],
\nll 
%10
{\cal H}_{\mp\mp\pm\mp} &=&
  \mp \frac{s k_4}{\srll}  \Biggl[
                           \cpl {\cal F}_{3}^{\pm}
                            - 2 {\cal F}_{5}^{\pm}
            +  \frac{1}{2} \cmi\left(
                                {\cal F}_{12}^{\pm}
          -  \frac{\kmi}{2} \cpl{\cal F}_{13}^{\pm} \right)
                                                  \Biggr],
\nll
%11
{\cal H}_{\mp\mp\pm\pm} &=&
 \pm \frac{s k_4}{\srll} \cpl \Biggl[
                             {\cal F}_{3}^{\pm}
               - \frac{1}{2} {\cal F}_{12}^{\pm}
       - \frac{\kmi}{4}  \cmi{\cal F}_{13}^{\pm}
                              \Biggr], 
\nll
%12
{\cal H}_{\pm\mp\mp\mp} &=&-\sqrt{\frac{s}{2}} k_3\Biggl[
                 \frac{2}{\kmi} {\cal F}_{0}^{\pm}
             +\cpl \left(
           \frac{\kmi}{4}   \cmi{\cal F}_{8}^{\pm}
                              + {\cal F}_{10}^{\pm} \right) \Biggr].
\eqa
Here the coefficients $k_{3,4,\pm}$ are defined by
\bqa
   k_3=\Nmuu\cos\frac{\vartheta_e}{2},\qquad k_4=\Nmuu\sin\frac{\vartheta_e}{2},
   \qquad k_{\pm}=s\pm\mz^2,
\eqa
with 
\bqa
   \Nmuu=\sqrt{\frac{s-\mz^2}{2} }\,,
\eqa
%---
$Z_2(\mel)$ and $Z_3(\mel)$ are the denominators of fermionic propagators:
\bqa
   && Z_2(\mel)=s-\mel^2\,,
   \nll
   && Z_3(\mel)=\frac{Z_2(\mel)}{2s}\left[s+\mel^2-\mz^2
   +\sqrt{\lambda(s,\mel^2,\mz^2)}\cos\vartheta_{e}\right],
\eqa
and $\vartheta_{e}$ denotes the $e^{\pm}$ scattering angle.
The Mandelstam variables transform as follows: 
\bqa
   && s \rightarrow
 -\frac{1}{2}\left[\left(s-\frac{\mz^2\mel^2}{s}-\mz^2-2\mel^2+\frac{\mel^4}{s}\right)
   -\frac{s-\mel^2}{s}\sqrt{\lambda(s,\mel^2,\mz^2)}\cos\vartheta_{e}\right],
   \nll
   && u \rightarrow
 -\frac{1}{2}\left[\left(s+\frac{\mz^2\mel^2}{s}-\mz^2-2\mel^2-\frac{\mel^4}{s}\right)
   +\frac{s-\mel^2}{s}\sqrt{\lambda(s,\mel^2,\mz^2)}\cos\vartheta_{e}\right],
   \nll
   && t \rightarrow s. \nonumber
\eqa

\section{Numerical results and comparison\label{Num_results_comp}}
%-----------------------------------------------------------------
In this section we present the \sanc predictions for various observables of all three 
processes under consideration. The tree level and single real photon emission
contributions are compared with {\tt CompHEP}, while one-loop electroweak
and QED corrections for the production channel $e\gamma\to e Z$ are checked against
the {Grace-loop} package~\cite{Belanger:2003sd} and Ref.~\cite{Denner:1992qf}.
Note that all numerical results of this section are produced with the
Standard \sanc INPUT (section 6.2.3 of Ref.~\cite{Bardin:2005dp})
if not stated otherwise.

\subsection{Annihilation channel $\bar{f}_1f_1\to Z\gamma$}
%----------------------------------------------------------
For this process we show  in Table~\ref{comparisoneeZA} a comparison
between \sanc and {\tt CompHEP} results for the Born level cross
sections and the cross sections of hard photon radiation.

\begin{table}[!ht]
\begin{center}
\begin{tabular} {||l|l|l|l|l|l||}
\hline
\hline
&\multicolumn{5}{|c||}{$\sigma$, pb}
\\
\hline
$\sqrt{s}$, GeV     & 100       & 200       & 500      & 1000      & 2000 \\
\hline
Born (\sanc)        & 2482.0(1) & 86.230(1) & 11.652(1)& 2.9845(1) & 0.77816(1) \\
Born ({\tt CompHEP})& 2482.0(1) & 86.230(1) & 11.651(1)& 2.9846(1) & 0.77817(1) \\
\hline
Hard (\sanc)        & 586.7(7)  & 43.26(8)  &  7.69(2) & 2.341(6)  & 0.717(2)   \\
Hard ({\tt CompHEP})& 586.7(3)  & 42.48(5)  &  7.47(1) &\mbox{unstable}&\mbox{unstable} \\
\hline
\hline
\end{tabular}
\end{center}
\caption{Comparison of the Born and Hard cross sections of the $e^+e^-\to Z\gamma(\gamma)$ 
         process ({\tt CompHEP} input, $E_{\gamma} \ge 1$ GeV).
         The uncertainty of the last significant digit is given in brackets.
         \label{comparisoneeZA}}
\end{table}

\begin{table}[!ht]
\begin{center}
\begin{tabular} {||l|l|l|l|l|l||}
\hline
\hline
&\multicolumn{5}{|c||}{$\sigma$, pb}
\\
\hline
$\sqrt{s}$, GeV     & 100       & 200       & 500       & 1000      & 2000       \\
\hline
Born (\sanc)        & 1349.5(1) & 49.086(1) & 6.9785(1) & 1.8469(1) & 0.49555(1) \\
Born ({\tt CompHEP})& 1349.4(1) & 49.086(1) & 6.9786(1) & 1.8469(1) & 0.49555(1) \\
\hline
Hard (\sanc)        & 173.82(3) & 14.138(3) & 2.7978(9) & 0.9228(4) & 0.3024(2)  \\
Hard ({\tt CompHEP})& 173.82(3) & 14.083(21)& 2.7627(21)& 0.9045(11)& 0.2936(4)  \\
\hline
\hline
\end{tabular}
\end{center}
\caption{The same as Table~\ref{comparisoneeZA} but for the process 
         $\mu^+\mu^-\to Z\gamma(\gamma)$.\label{comparisonmmZA}}
\end{table}

As can be seen from Table~\ref{comparisoneeZA}, we found very good agreement
for the Born cross section. For the hard contribution we have perfect
agreement at $\sqrt{s}=$100 GeV, then a difference rapidly rising with energy,
and eventually unstable {\tt CompHEP} predictions for $\sqrt{s}$ at and above 1 TeV. 
As seen from Table~\ref{comparisonmmZA}
for the process $\mu^+\mu^-\to Z\gamma(\gamma)$, the hard contributions
stay closer (though statistically incompatible) within a wider range of $\sqrt{s}$
pointing to the origin of the difference due to collinear singularities of the integrand.
The stability against variation of $\bar{\omega}$ discussed below gives us a
great level of confidence in the \sanc results.

%------------------------
In Tables~\ref{sigee10M}--\ref{sigddZA10M} we present the results of
our calculations for the annihilation channels $e^{+}e^{-}\to Z\gamma(\gamma)$,
$\bar{u}u\to Z\gamma (\gamma)$ and $\bar{d}d\to Z\gamma (\gamma)$, respectively,
carried out with 10M statistics for the hard cross section
for five energies and at each energy for 
two values of $\bar\omega$: $\bar\omega = 10^{-5}\sqrt{s}/2$ (subscript 1) and
$10^{-6}\sqrt{s}/2$ (subscript 2); for $\sigma$'s in pb and for
$\delta=\sigma^{\rm 1-loop}/\sigma^{\rm Born}-1$ in \%.

%------------------------
% Note: this table is based on the computer output of 09-Oct-2007
%------------------------
\begin{table}[!h]
\begin{center}
\begin{tabular} {||l|l|l|l|l|l||}
\hline
\hline
$\sqrt{s}$, GeV               &    200   &    500    & 1000      & 2000       & 5000        \\
\hline
$\sigma^{\rm Born}$, pb       &27.8548(1)& 3.37334(1)&0.816485(2)& 0.202534(1)& 0.0323355(1) \\
\hline
$\sigma^{\rm 1-loop}_{1}$, pb &43.36(4)  & 5.216(9)  &1.239(4)   & 0.299(1)   & 0.0436(3)   \\
\hline
$\sigma^{\rm 1-loop}_{2}$, pb &43.38(5)  & 5.211(10) &1.235(4)   & 0.298(2)   & ---         \\
\hline
$\delta_{1}$, \%              &  55.7(2) &  54.6(3)  &  51.9(4)  &  47.4(6)   & 34.9(8)     \\
\hline
$\delta_{2}$, \%              &  55.7(2) &  54.5(3)  &  51.3(5)  &  46.9(8)   & ---         \\
\hline
\hline
\end{tabular}
\end{center}
\caption{Comparison of  the Born and one-loop cross sections of the annihilation channel
$e^{+}e^{-}\to Z\gamma (\gamma)$ calculated with different
values of the soft/hard separation parameter $\bar\omega$; for details see the text.}
%[[db: Table to be recomputed with improved IZi]]
\label{sigee10M}
\end{table}

%------------------------
% Note: this table is based on the computer output of 21-Sep-2007
%       the values for 5000 GeV are from 24-Sep-2007
%------------------------
\begin{table}[!h]
\begin{center}
\begin{tabular} {||l|l|l|l|l|l||}
\hline
\hline
$\sqrt{s}$, GeV               &    200   &    500    & 1000       & 2000       & 5000         \\
\hline
$\sigma^{\rm Born}$, pb       & 4.7504(1)& 0.57540(0)& 0.13927(0) & 0.034548(0)& 0.005516(0) \\
\hline
$\sigma^{\rm 1-loop}_{1}$, pb & 5.3399(8)& 0.6472(2) & 0.15367(6) & 0.036458(2)& 0.005203(6) \\
\hline
$\sigma^{\rm 1-loop}_{2}$, pb & 5.3392(9)& 0.6470(2) & 0.17159(7) & 0.036458(2)& 0.005193(7) \\
\hline
$\delta_{1}$, \%              & 12.41(2) & 12.48(3)  & 10.34(4)   &  5.54(6)   & -5.67(11)   \\
\hline
$\delta_{2}$, \%              & 12.39(2) & 12.44(3)  & 10.28(5)   &  5.53(8)   & -5.84(12)   \\
\hline
\hline
\end{tabular}
\end{center}
\caption{Comparison of the Born and one-loop cross sections of the annihilation channel
$\bar{u}u\to Z\gamma (\gamma)$ calculated with different
values of the soft/hard separation parameter $\bar\omega$.}
\label{siguu10M}
\end{table}

%------------------
%%          Channel dd2ZA(A) *** results of 24-Sep-2007         
%==========================================================================

\begin{table}[!ht]
\begin{center}
\begin{tabular} {||l|l|l|l|l|l||}
\hline
\hline
$\sqrt{s}$, GeV               &    200   &   500     & 1000       & 2000       & 5000        \\
\hline
$\sigma^{\rm Born}$, pb       & 1.5230(0)& 0.18450(0)& 0.044658(0)& 0.011078(0)& 0.0017686(0)\\
\hline
$\sigma^{\rm 1-loop}_1$, pb   & 1.6033(1)& 0.18920(1)& 0.043823(4)& 0.009992(2)& 0.0012825(4)\\
\hline
$\sigma^{\rm 1-loop}_2$, pb   & 1.6033(1)& 0.18924(1)& 0.043825(5)& 0.009992(2)& 0.0012826(4)\\
\hline
$\delta_1$, \%                & 5.274(4) & 2.549(6)  & -1.869(10) &  -9.807(14)& -27.486(23) \\
\hline
$\delta_2$, \%                & 5.275(4) & 2.570(7)  & -1.865(11) &  -9.804(16)& -27.479(27) \\
\hline
\hline
\end{tabular}
\end{center}
\caption{Comparison of the Born and one-loop cross sections of the annihilation channel
$\bar{d}d\to Z\gamma (\gamma)$ calculated with different
values of the soft/hard separation parameter $\bar\omega$.}
\label{sigddZA10M}
\end{table}

The total 1-loop cross section $\sigma^{\rm 1-loop}$ is the sum of the Born, virtual, soft and hard
contributions:
$$
 \sigma^{\rm 1-loop} = \sigma^{\rm Born}
+\sigma^{\rm virtual}(\lambda)+\sigma^{\rm soft}(\lambda,\bar\omega)
+\sigma^{\rm hard}(\bar\omega).
$$
Here $\sigma^{\rm virtual}$ and $\sigma^{\rm soft}$ depend on the regularizing
parameter $\lambda$ which cancels in their sum. This cancellation was
checked on the algebraic level. The contributions
 $\sigma^{\rm soft}$ and $\sigma^{\rm hard}$
depend on $\bar\omega$, the soft/hard separation parameter. This dependence
must cancel on the numerical level. To ascertain this cancellation we have
done the calculation at each energy for two values of $\bar\omega$ as shown
above. Comparing the corresponding values of $\sigma^{\rm 1-loop}$ and of
$\delta$ we can see that there is no change outside the statistical errors of the Monte Carlo 
integration.

The following cuts were imposed:
\begin{itemize}
\item
CMS angular cuts for the Born, soft and virtual contributions where
there is only one photon in the final state:
$
\vartheta_{\gamma,\,Z}\in[1^\circ,179^\circ]
$
\item
CMS angular cuts on the $Z$ boson and on the two photons and CMS energy
cuts on the photons for the hard contribution: for the event to be accepted,
$\vartheta_Z$ and at least one of $\vartheta_{\gamma_1}$ or $\vartheta_{\gamma_2}$
must lie in the interval $[1^\circ,179^\circ]$, and both
photons must have a CMS energy greater than $\bar\omega$.
\end{itemize}

For all tables, the numbers in brackets give the statistical uncertainties of the last
digit shown.

\subsection{Decay channel $Z\to f_1\bar{f}_1\gamma$}
%-------------------------------------------
 In Table~\ref{comparisoneZeeA} we present the results of a comparison of 
the Born cross section and the cross section of hard photon bremsstrahlung of $Z$
 boson decay between 
\sanc and {\tt CompHEP}. We see that we have excellent agreement between these
 two programs. Differences are within statistical errors.

\begin{table}[!ht]
\begin{center}
\begin{tabular} {||l|l|l|l|l||}
\hline
\hline
&\multicolumn{4}{|c||}{$\Gamma$, GeV}
\\
\hline
${\omega}$, GeV & 0.1          & 1            & 2             & 5
\\
\hline
Born (SANC)         & 0.027730(1) & 0.015779(1) & 0.012269(1) & 0.0078271(1) \\
Born (CompHEP)      & 0.027730(1) & 0.015778(1) & 0.012269(1) & 0.0078268(1) \\
\hline
Hard (SANC)         & 0.004393(2) & 0.001358(1) & 0.0007944(4)& 0.0002941(2) \\
Hard (CompHEP)      & 0.004392(3) & 0.001359(1) & 0.0007940(5)& 0.0002946(2) \\
\hline
\hline
\end{tabular}
\end{center}
\caption{Comparison of the Born and Hard widths
of the $Z \to \mu^+ \mu^- \gamma (\gamma)$ decay (CompHEP input,
$E_{\gamma} > {\omega}$ for photon(s)).
The uncertainty of the last significant digit is given in brackets.}
\label{comparisoneZeeA}
\end{table}

In Table~\ref{table:Z2mumuAA} we show the differential decay rate
$d\Gamma/ds\times 10^8$ in GeV$^{-1}$ of the decay $Z\to \mu^+\mu^-\gamma(\gamma)$,
where $\sqrt{s}$ is the invariant mass of the $\mu^+\mu^-$ pair
calculated with two different values of the soft/hard separation
parameter $\bar\omega$: $10^{-4}$ GeV (subscript 1) and $10^{-5}$ GeV (subscript 2).
The quantity  $\delta$ is given by
$\delta = (d\Gamma^{\rm 1-loop}/ds-d\Gamma^{\rm Born}/ds)/d\Gamma^{\rm Born}/ds$.

\begin{table}[!ht]
\begin{center}
\begin{tabular}{||c|l|l|l|l|l|l||}
\hline
\hline
 $\sqrt{s}$, GeV                  & 1.        & 10.       &   20.      & 50.     & 70.     \\
\hline
$d\Gamma^{\rm Born}/ds$, GeV$^{-1}$ & 7.918   & 18.57     & 22.63      & 39.98   & 89.11   \\
\hline
$d\Gamma_1^{\rm 1-loop}/ds$, GeV$^{-1}$& 744.21(4)& 18.834(4)& 21.949(8) & 35.92(2)& 76.12(6)\\
$d\Gamma_2^{\rm 1-loop}/ds$, GeV$^{-1}$& 744.21(4)& 18.830(5)& 21.937(10)& 35.93(3)& 76.16(8)\\
\hline
$\delta_1$                  & 92.992(5) & 0.0140(2) & -0.0300(3) & -0.1014(6) & -0.1458(7) \\
$\delta_2$                  & 92.992(5) & 0.0137(3) & -0.0305(4) & -0.1014(8) & -0.1452(9) \\
\hline
\hline
\end{tabular}
\caption{Comparison of the Born and one-loop differential widths of the decay channel 
$Z\to \mu^+\mu^-\gamma(\gamma)$ calculated with different values of the soft/hard separation 
parameter $\bar\omega$; for details see the text.}
\label{table:Z2mumuAA}
\end{center}
\end{table}

\begin{figure}[!h]
\begin{center}
\includegraphics[width=12cm,height=10cm]{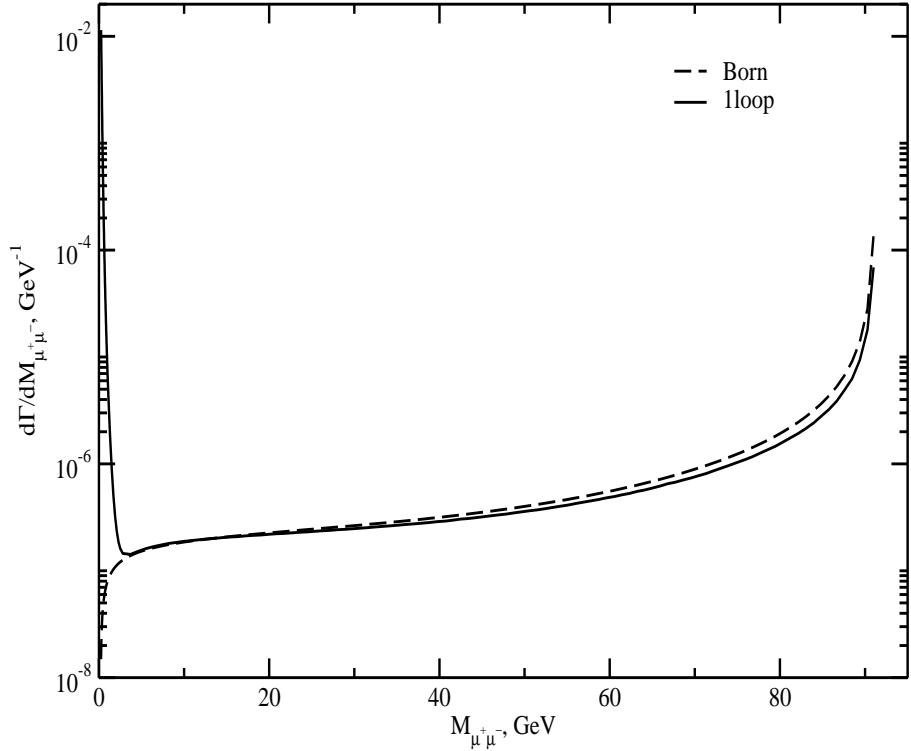}
\caption{Invariant mass distribution of the $\mu^+\mu^-$ pair for the decay 
$Z\to\mu^+\mu^-\gamma(\gamma)$. Both the Born (dashed line) and 
the 1-loop (dotted line) results are shown.}
%$d\Gamma^{\rm 1-loop}/ds$ 
\label{dGamma_ds}
\end{center}
\end{figure}

In Fig.~\ref{dGamma_ds} we show the differential decay widths $d\Gamma^{\rm Born}/ds$ and 
$d\Gamma^{\rm 1-loop}/ds$ for the decay $Z\to \mu^+\mu^-\gamma(\gamma)$ as functions of
$\sqrt{s}=M_{\mu{+}\mu^{-}}$.

%\vspace*{16mm}

%-----------
%[[db: Figure will be rebuilt.]

The Coulomb peak, which is due to photon exchange in the Feynman one-loop diagram  
with a $\gamma Z\gamma$ three-boson vertex, is clearly seen.

\clearpage

\subsection{$Z$ production channel $e\gamma\to e Z$}
%---------------------------------------------------
 As can be seen from Table~\ref{comparisoneAeZe}, we have again very good agreement 
between \sanc and {\tt CompHEP} predictions for the tree level and real photon 
emission cross sections of this process.

\begin{table}[h]
\begin{center}
\begin{tabular} {||l|l|l|l|l|l||}
\hline
\hline
&\multicolumn{5}{|c||}{$\sigma$, pb}
\\
\hline
$\sqrt{s}$, GeV      & 100       & 200       & 500       & 1000      & 2000       \\
\hline
Born (\sanc)         & 82.266(1) & 23.716(1) & 5.5747(1) & 1.5343(1) & 0.40648(1) \\
Born ({\tt CompHEP}) & 82.265(1) & 23.716(1) & 5.5747(1) & 1.5343(1) & 0.40647(1) \\
\hline
Hard (\sanc)         &  4.012(1) &  3.689(2) &  1.368(1) & 0.4986(6) & 0.1682(3)  \\
Hard ({\tt CompHEP}) &  4.014(0) &  3.688(1) &  1.364(1) & 0.4973(6) & 0.1678(3)  \\
\hline
\hline
\end{tabular}
\end{center}
\caption{Comparison of the Born cross sections
        for the $\gamma e^{-}\to Ze^{-}$ reaction
        and of the Hard cross sections for the $\gamma\mu^{-}\to
Z\mu^{-}\gamma$ reaction
        ({\tt CompHEP} input, $E_{\gamma} \ge 1$ GeV).
        }
\label{comparisoneAeZe}
\end{table}
%----------------------------

In Table~\ref{sig_eg2eZ} we present the results of our calculations for
the channel $\gamma e^-\to Ze^-(\gamma)$ carried out with
10M statistics for the hard cross section for five energies and at each energy 
for two values of $\bar\omega$: $\bar\omega = 10^{-5}\sqrt{s}/2$ (subscript 1) and
$10^{-6}\sqrt{s}/2$ (subscript 2). 

\begin{table}[!ht]
\begin{center}
\begin{tabular} {||l|l|l|l|l|l||}
\hline
\hline
 $\sqrt{s}$, GeV              &    200  &    500   &   1000   &   2000   & 5000 \\
\hline
 $\sigma^{\rm Born}$, pb      &8.3381(3)&1.79168(0)&0.46840(0)&0.11842(0)&0.019007(0)\\
\hline
$\sigma^{\rm 1-loop}_{1}$, pb &8.7988(5)&1.9591(2)&0.52129(5)&0.13171(1)&0.02037(2)\\
\hline
$\sigma^{\rm 1-loop}_{2}$, pb &8.8002(9)&1.9593(2)&0.52131(6)&0.13168(1)&0.02037(3)\\
\hline
$\delta_{1}$, \%              & 5.54(1) & 9.35(1) & 11.29(1) & 11.23(1) &7.16(1) \\
\hline
$\delta_{2}$, \%              & 5.54(1) & 9.36(1) & 11.30(1) & 11.20(1) &7.15(2) \\
\hline
\hline
\end{tabular}
\end{center}
\caption{Comparison of the Born and 1-loop cross sections of channel
$\gamma e^-\to Ze^-(\gamma)$ calculated with different
values of the soft/hard separation parameter $\bar\omega$.}
\label{sig_eg2eZ}
\end{table}

The notation of the various contributions, $\sigma^{\rm Born}$ etc.,
is as in the previous case.

The cancellation of the $\lambda$-dependent terms was again checked
on the algebraic level.
The cancellation of the $\bar\omega$ dependence on the numerical level
was tested as in the previous case.
Comparing the corresponding values of $\sigma^{\rm 1-loop}$ and of $\delta$
we can see again that there is no change outside the statistical errors of the Monte
Carlo integration.

The following cuts were imposed:
\begin{itemize}
\item
CMS angular cuts for the Born cross section and for the contributions
with Born-like kinematics:
$
\vartheta_{e,\,Z}\in[1^\circ,179^\circ]
$
\item
CMS angular cuts on the $Z$ boson and on the photon and a CMS energy
cut on the electron for the hard contribution: for the event to be accepted,
$\vartheta_Z$ and $\vartheta_{e}$
must lie in the interval $[1^\circ,179^\circ]$, and the
photon must have a CMS energy greater than $\bar\omega$.
\end{itemize}

The numbers in brackets give the statistical uncertainties of the last
digit shown.

\clearpage
%---------------------------

 In Table~\ref{comparisoneAeZe_denner} we show the comparison of the Born
cross sections: the angular distributions $d\sigma/d\cos\vartheta_{e}$
and the cross sections integrated over the given angular intervals,
as well as the 1-loop EW corrections $\delta$,
produced by three programs: that of Ref.~\cite{Denner:1992qf},
{Grace-loop}~\cite{Belanger:2003sd} and \sanc.

%\vspace*{-1.5cm}
\begin{table}[!ht]
\begin{center}
\begin{tabular} {||c|c|c|r|r|r||}
\hline
\hline
$\sqrt{s}$, GeV &$\vartheta$&&\multicolumn{1}{c}{ Ref.~\cite{Denner:1992qf}}
&\multicolumn{1}{|c|}{Grace-loop}&\multicolumn{1}{c||}\sanc\\
\hline
                &$20^{\circ}$ & $\sigma^{\rm Born}$, pb & 0.3931  &         & 0.39308  \\
                &             & $\delta$, \%            &  -5.96  &         & -5.9556  \\
                &$90^{\circ}$ & $\sigma^{\rm Born}$, pb & 0.6491  &         & 0.64906  \\
                &             & $\delta$, \%            &  -8.56  &         & -8.5562  \\
100             &$160^{\circ}$& $\sigma^{\rm Born}$, pb &  9.038  &         &  9.0383  \\
                &             & $\delta$, \%            & -10.00  &         & -10.005  \\
                &\aga         & $\sigma^{\rm Born}$, pb & 13.051  &         & 13.051   \\
                &             & $\delta$, \%            &  -9.04  &         & -9.0389  \\
                &\agb         & $\sigma^{\rm Born}$, pb & 33.484  &         & 33.484   \\
                &             & $\delta$, \%            & -10.27  &         & -10.273  \\
\hline
                &$20^{\circ}$ & $\sigma^{\rm Born}$, pb & 0.02898 &         & 0.028984 \\
                &             & $\delta$, \%            &  -30.08 &         &  -30.079 \\
                &$90^{\circ}$ & $\sigma^{\rm Born}$, pb & 0.03598 &         & 0.035985 \\
                &             & $\delta$, \%            &  -26.74 &         &  -26.744 \\
500             &$160^{\circ}$& $\sigma^{\rm Born}$, pb &  0.4661 &         &  0.46607 \\
                &             & $\delta$, \%            &  -23.05 &         &  -23.054 \\
                &\aga         & $\sigma^{\rm Born}$, pb &  0.7051 & 0.70515 &  0.70515 \\
                &             & $\delta$, \%            &  -25.69 & -25.689 &  -25.690 \\
                &\agb         & $\sigma^{\rm Born}$, pb &   1.770 &  1.7696 &   1.7697 \\
                &             & $\delta$, \%            &  -22.31 & -22.313 &  -22.313 \\
\hline
                &$20^{\circ}$ & $\sigma^{\rm Born}$, pb & 0.001869&         & 0.0018688\\
                &             & $\delta$, \%            &  -41.57 &         &  -41.575 \\
                &$90^{\circ}$ & $\sigma^{\rm Born}$, pb & 0.002334&         & 0.0023340\\
                &             & $\delta$, \%            &  -41.98 &         &  -41.981 \\
2000            &$160^{\circ}$& $\sigma^{\rm Born}$, pb &  0.03094&         &  0.030942\\
                &             & $\delta$, \%            &  -33.99 &         &   -33.994\\
                &\aga         & $\sigma^{\rm Born}$, pb &  0.04620& 0.046201&  0.046201\\
                &             & $\delta$, \%            &  -39.53 & -39.529 &   -39.529\\
                &\agb         & $\sigma^{\rm Born}$, pb &  0.1170 &  0.1170 &  0.11697 \\
                &             & $\delta$, \%            &  -30.84 & -30.845 &  -30.845 \\
\hline\hline
\end{tabular}
\end{center}
\vspace*{-3mm}
%[[WvS: note a slight change of wording in the caption:]]
\caption{Triple comparison of the Born cross section and of the correction
 $\delta = \sigma^{\rm 1-loop}/\sigma^{\rm Born}-1$
 for channel $\gamma e^- \to Z e^-(\gamma)$ 
(Ref.~\cite{Denner:1992qf} input, $E_\gamma = 0.025 \sqrt{s}$\,GeV).}
\label{comparisoneAeZe_denner}
\end{table}

We have excellent agreement between these three results.
Note that in this table the results taken from the literature were given there without
statistical errors. The statistical errors of numbers obtained with \sanc are
in the digits beyond those shown.

\section{Conclusions\label{concl}}
%--------------------------------- 
In this paper we describe the implementation of the complete one-loop EW
calculations, including hard bremsstrahlung contributions, for the process 
$f_1\bar{f}_1 ZA \to 0$ into the \sanc framework.  The calculations were
done using a combination of analytic and Monte Carlo integration methods which
make it easy to calculate a variety of observables and to impose experimental cuts.
We have presented analytical expressions for the covariant amplitudes of the process and 
the helicity amplitudes for three different cross channels: 
$Z$ boson production $f_1\bar{f}_1\to Z\gamma$ and $f_1\gamma\to f_1 Z$, and
for the decay $Z\to f_1\bar{f}_1 \gamma$.
To be assured of the correctness of our analytical results, we
observe the independence of the form factors on gauge parameters 
(all calculations were done in $R_{\xi}$ gauge),
the validity of the Ward identity for the covariant amplitudes.
We have compared our numerical results for these processes with other independent calculations.
The Born level and the hard photon contrubutions of all three channels were checked against 
{\tt CompHEP} package and we found a very good agreement 
except for the annihilation channel at high energies.
For the channel $\gamma e^-\to Z e^-(\gamma)$, the comparison of
 the \sanc  EW NLO predictions 
with the results of Refs.~\cite{Denner:1992qf,Belanger:2003sd} has shown an
excellent agreement in a wide range of CMS energies and final electron scattering angles.

The results presented lay a base for subsequent extensions of calculations in the 
annihilation channel appropriate to the process $pp\to X Z\gamma$ at hadron colliders.

\vspace*{2cm}
{\bf Acknowledgements.}
%\begin{acknowledgement}
%       acknowledgment acknowledgment acknowledgment 
This work is partly supported by INTAS grant $N^{o}$ 03-51-4007 \\ 
and by the EU grant mTkd-CT-2004-510126 in partnership with the
CERN Physics Department and by the Polish Ministry of Scientific Research and
Information Technology grant No 620/E-77/6.PRUE/DIE 188/2005-2008 and
by Russian Foundation for Basic Research grant $N^{o}$ 07-02-00932.

WvS is indebted to the directorate of 
the Dzhelepov Laboratory of Nuclear Problems, JINR, Dubna
for the hospitality extended to him during September 2007.
%\end{acknowledgement}

\newpage
\bibliographystyle{utphys_spires}
\addcontentsline{toc}{section}{\refname}\bibliography{FFZA}

\end{document}